# Sound Transmission Through a Finite-Sized Double Panel Cavity with a Micro-Perforated Panel Insertion

Zhenwei. Zhou [a,*], Jiaming. Wu [a,**], and Zhiyu. Yang [b,***]

[a] School of Civil Engineering and Transportation, South China University of Technology, Guangzhou, 510641, Guangdong, China

[b] Department of Physics, Hong Kong University of Science and Technology, Clear Water Bay, Kowloon, Hong Kong, China.

*e-mail: zw.zhou@hotmail.com

**e-mail: ctjmwu@scut.edu.cn

***e-mail: phyang@ust.hk

**Abstract-** This paper proposes a noise insulation cavity composed of two parallel plates and a micro-perforated plate insertion parallel to the plates, which divides the cavity between the plates into two parts. A theoretical model was established that takes into account of all the couplings among the major parts of the structure, namely the two solid plates, the perforated plate, and the air cavity, together with the simply support boundary conditions. Numerical calculations were performed with different parameters of the micro-perforated plate including its position, perforation ratio, plate thickness, and hole diameters. The calculations indicated that the proposed double-panel structure with a micro-perforated plate insertion exhibited significant improvements in the sound transmission loss (STL) in certain frequency range as compared to a double- or triple-panel structure without a micro-perforated plate. Below 200 Hz the improvement in STL is mainly due to the weakening of the resonances by the energy dissipation of the perforated plate, while in the medium to high frequency range the STL enhancement is mostly due to the dissipation by the perforated plate in the broad frequency band. The theoretical results are in good agreement with the experimental results.

*Key words:* Micro-perforated panel; Sound insulation; Double panel

## 1. INTRODUCTION

A double-panel, or double-leaf structure is an air layer sealed between two plates. Given the high sound transmission loss (STL) characteristics, the structure is widely used in aviation, automobiles, trains, machinery, etc.(Brunskog, 2005; Chazot & Guyader, 2007; Díaz-Cereceda, Poblet-Puig, & Rodríguez-Ferran, 2012; J. Wang, Lu, Woodhouse, Langley, & Evans, 2005; Xin & Lu, 2009, 2011a; Xin, Lu, & Chen, 2010). Brunskog (Brunskog, 2005) proposed a model that considered the stud effect to examine a double-leaf structure with double-headed bolts. Xin et al. (Xin & Lu, 2009, 2011a; Xin et al., 2010) analyzed the analytical equations of double-plate with a simple support and under fixed support conditions to calculate their STL. Chazot et al. (Chazot & Guyader, 2007) proposed a patch-mobility method to predict the STL for a finite-sized double-plate coupled with an air cavity which significantly improved the computational efficiency. Although previous studies indicate that double-layer plates provide a better noise barrier when compared with a single-layer plate, the STL of a double-layer panel at the resonant frequency is significantly reduced owing to the effect of the "mass–air–mass" resonance, leading to poor STL at low frequency. To solve the problem, extensive studies propose sandwiching various materials in between the double-plates to replace the air cavity and transform the double-plates into a multi-layered sandwich structure (Chazot & Guyader, 2009; Guyader, Cacciolati, & Chazot, 2010; Panneton & Atalla, 1996; Sgard, Atalla, & Amedin, 2007; Sun & Liu, 2016; Xin & Lu, 2011b; Zhou, Bhaskar, & Zhang, 2013b). Bolton et al. (Bolton & Green, 1993; Bolton, Shiau, & Kang, 1996) proposed an elastic porous material lined with double layer plates and used Biot's theory to calculate the STL of the structure. Based on Biot's theory, Liu et al. (Liu & Daudin, 2017; Liu & He, 2016) outlined a theoretical calculation model for elastic porous materials sandwiched between finite-sized plates under a fixed-support boundary condition. Zhou et al. (Zhou, Bhaskar, & Zhang, 2013a) calculated and optimized the STL of an infinitely extended double-plate plus an elastic porous insertion. Chazot et al. (Chazot & Guyader, 2009) applied the patch-mobility method to calculate the STL of a double-plate sandwiched between elastic porous materials. The multi-layer plate structure with an elastic porous material in the cavity significantly improved the STL at the resonant frequency. Although the successful applications of sandwiched panels are known for solving the STL problem, the porous material fibers can irritate the eyes and skin and cause various respiratory ailments. To deal with the undesired health effects, it is necessary to develop a non-fiber noise reduction device, such as micro-perforated plates.

A micro-perforated plate is a good sound-absorbing device. It was first proposed by Maa in 1975 (Maa, 1975). The study indicated that when a micro-perforated plate was placed at a certain distance in front of a rigid plate, a cavity was formed between them. Thus, the air in the micro-hole could be considered as a mass, the cavity behind the hole could serve as an air spring, and the complete structure corresponded to a resonant absorber similar to a Helmholtz resonator. The rigid wall plate at the back provide noise insulation, and it is extremely effective in improving the acoustic comfort and speech intelligibility in enclosed space, such as theatrical halls. It is simple in structure, resistant to corrosion, and can also satisfy light, transparent, and fiber-free requirements. Therefore, micro-perforated plates are being increasingly used in building acoustics, aviation, and ground transportation industries. Currently, several studies are focusing on sound absorbing structures with rigid walls behind the micro-perforated plates. The results indicate that when the perforation rate of the micro-perforated plates is between 0.5% and 1.5% and the diameter of the small holes is between 0.05 and 1 mm, the micro-perforated plate exhibits extremely good sound absorption effect. The sound absorption bandwidth of a micro-perforated plate absorber is typically narrow (generally 1 to 2 octaves). A multi-layer micro-perforated board sound-absorbing structure was proposed to widen the effective sound absorption bandwidth of the sound-absorbing structure (Bravo, Maury, & Pinhède, 2013; Omrani & Tawfiq, 2011; Sakagami, Matsutani, & Morimoto, 2010). In this structure, the single-layer micro-perforated plate generally exhibited only a single resonance absorption peak, and significantly improved the sound absorption performance. However, the large thickness of the multilayer structure made it difficult to install it in a space-constrained environment. Further in-depth research were conducted on micro-perforated plates and ultra-micro-perforated plates (Qian, Kong, Liu, Sun, & Zhao, 2013), irregular structured back cavities (C. Yang, Cheng, & Pan, 2013), micro-perforated plate with Helmholtz resonators (Gai, Xing, Li, Zhang, & Wang, 2016), and deep back cavities (C. Wang & Huang, 2011). However, all these reports focus on increasing the sound absorption coefficient of the micro-perforated plate structure, broadening the frequency of the sound absorption, perfecting the theory of micro-perforated plates, and significantly assisting in practical applications. Most existing studies on micro-perforated plates involve infinitely rigid plates. A micro-perforated plate is generally a thin plate in which its own vibration is coupled to the entire structure; therefore, it is difficult to ignore the resulting effects. Extensive studies on micro-perforated plates mainly concentrate on the sound absorption coefficient of the micro-perforated plates, but there are few studies on their sound insulation performance. Putra et al. (Putra, Ismail, & Ayob, 2013) examined the sound insulation performance of a three-layer structure composed of a micro-perforated plate inserted in an infinitely extended double-plate structure. Maury et al. (Maury, Bravo, & Pinhede, 2013) investigated the micro-perforated plate model of a flexible back plate by considering the vibration coupling of the plate and sound insulation properties of the double-layer structure. While these studies provided some insight into the sound insulation performance improvement of the micro-perforated plate, a more realistic model that takes into account the finite size of the plates and their boundary conditions, the vibration of not only the plates but also the perforated insertion under incident sound field, and the full coupling among the upper and lower plates and the perforated insertion mediated by the air cavities, is still absent.

This study deals with the STL of a sound blocking structure composed of a finite-sized double-plate with micro-perforated plates (DPMPP) inserted in between. The above-mentioned aspects are all included in the theoretical model. The paper is divided in the following sections. Section 0 introduces the theoretical model for calculating the STL of the structure. Section 3 describes the experiment for validating the theoretical model. Based on the coupling model proposed in Section 0, Section 4 compares the effect of the micro-perforated plate parameters on the sound insulation performance.

## 2. THEORY

Figure 1 shows the system under investigation. A triple-panel partition with three parallel panels ($L_x \times L_y$) and two air cavities ($L_x \times L_y \times h$) is surrounded by a baffled wall. The panel at the top is called the incidence panel and that at the bottom is referred to as the radiation panel. The middle panel is a micro-perforated panel (MPP). The triple-panel partition is subjected to a dynamic force via an incident pressure ($p_{inc}$) on the incidence panel. The distance between the incidence panel and MPP is $z_1$, the distance between MPP and the radiation panel is $z_2$, and the distance between the incidence panel and the radiation panel is $z_3$. It is assumed that normal displacement of the incidence panel $w^{IN}(x,y,t)$, the radiation panel $w^{RE}(x,y,t)$, and the MPP $w^{MPP}(x,y,t)$ are positive toward the positive z-axis. The harmonic time dependence is assumed throughout the study as $e^{i\omega t}$. Superscripts I, II, IN, MPP, and RE refer to cavity I, cavity II, the incidence panel, the MPP, and the radiation panel, respectively. The equations for the bending motion of the panels subjected to incident wave $p_{inc}(x,y,z^{IN},t)$ are expressed as follows:

For the incidence panel:
$$D^{IN}\nabla^4 w^{IN}(x,y,t) + m^{IN}\ddot{w}^{IN}(x,y,t) = 2p_{inc}(x,y,z^{IN},t) - p^I(x,y,z^{IN},t), \quad (1)$$

For the MPP:
$$D^{MPP}\nabla^4 w^{MPP}(x,y,t) + m^{MPP}\ddot{w}^{MPP}(x,y,t) = p^I(x,y,z^{MPP},t) - p^{II}(x,y,z^{MPP},t), \quad (2)$$

For the radiation panel:
$$D^{RE}\nabla^4 w^{RE}(x,y,t) + m^{RE}\ddot{w}^{RE}(x,y,t) = p^{II}(x,y,z^{RE},t), \quad (3)$$

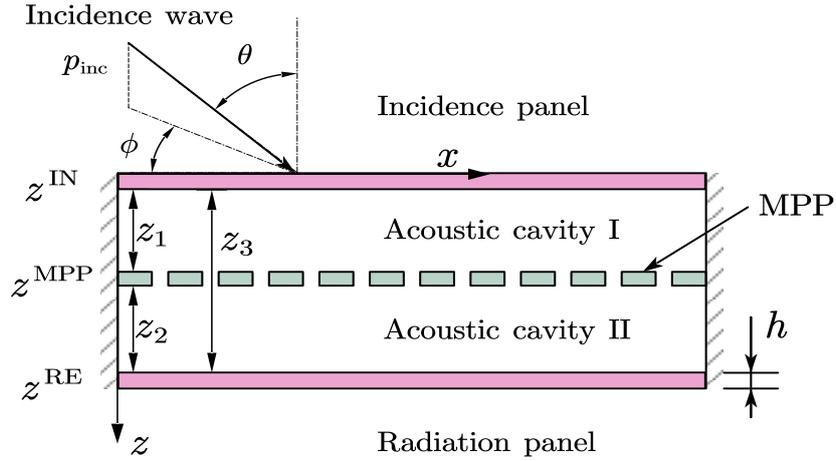

Figure 1 Baffled triple-panel partition with an MPP

where $m^{IN}$, $m^{RE}$, and $m^{MPP}$ denote the area mass density of the incidence panel, the radiation panel, and the MPP, respectively; $z^{IN}$, $z^{RE}$, and $z^{MPP}$ denote the scale on the $z$ axis of the incidence panel, the radiation panel, and the MPP, respectively; $D = h^3 E/12(1-\mu^2)$ denotes the flexural rigidity of the panel; $h$ denotes the thickness of the panel; $E$ is the Young's modulus; $\mu$ is the Poisson's ratio, and $p(x,y,z,t)$ is the pressure response in the cavity. In the above equations, the fluid loadings on the outside of both the panels are ignored, because it was found that fluid loading effects are absent even for the cases of a rubber membrane only 0.2 mm thick(Z. Yang, Mei, Yang, Chan, & Sheng, 2008).

The wave equation governing the pressure fields inside the air cavity is expressed as follows:

For cavity I:

$$\nabla^2 p^I(x,y,z,t) - \frac{1}{c^2}\ddot{p}^I(x,y,z,t) = -2\rho_0 \ddot{w}^{IN}(x,y,t)\delta(z-z^{IN}) + 2\rho_0 \dot{v}^{MPP}(x,y,t)\delta(z-z^{MPP}), \quad (4)$$

For the cavity II:

$$\nabla^2 p^{II}(x,y,z,t) - \frac{1}{c^2}\ddot{p}^{II}(x,y,z,t) = -2\rho_0 \dot{v}^{MPP}(x,y,t)\delta(z-z^{MPP}) + 2\rho_0 \ddot{w}^{RE}(x,y,t)\delta(z-z^{RE}), \quad (5)$$

where $\delta(x-x_0)$ denotes the Dirac delta function, $c$ denotes the speed of sound, $\rho_0$ denotes the air density, and $v^{MPP}(x,y,t)$ denotes the normal velocity of the MPP. It is assumed that the distance between the MPP holes is significantly shorter than the acoustic wavelengths, and the particle velocity at the MPP surface is modified by the fluid flow velocity $v_h$ through the holes. Net particle velocity $v^{MPP}$ formed by combining the normal velocity of the MPP $\tilde{v}^{MPP}$ and the fluid motion velocity $v_h$ is expressed as follows (Takahashi & Tanaka, 2002):

$$v^{MPP}(x,y,t) = (1-\sigma)\tilde{v}^{MPP}(x,y,t) + \sigma v_h(x,y,t), \quad (6)$$

where $\sigma$ denotes the perforation ratio, or the fraction area of the holes. Particle velocity $v_h$ is expressed in terms of the pressure difference of the cavities at its position and the acoustic impedance $Z^{MPP}$ as follows:

$$v_h(x,y,t) = \frac{p^I(x,y,z^{MPP},t) - p^{II}(x,y,z^{MPP},t)}{Z^{MPP}} \quad (7)$$

Acoustic impedance of the MPP $Z^{MPP}$ for a circular hole with diameter $d$ and length $t_h$ (thickness of the MPP) that are much smaller than the acoustic wavelength is given by Maa as follows (Maa, 1975):

$$Z^{MPP} = \frac{32\eta t_h}{d^2}\left(\sqrt{1+\frac{k_h^2}{32}}+\frac{\sqrt{2}}{8}k_h\frac{d}{t_h}\right)+i\omega\rho_0 t_h\left(1+\frac{1}{\sqrt{9+k_h^2/2}}+\frac{8}{3\pi}\frac{d}{t_h}\right) \tag{8}$$

where $\eta$ denotes the viscosity coefficient of air, and $k_h$ denotes the ratio that is $\sqrt{2}$ times the small-hole radius ($\sqrt{2}d/2$) over the viscous boundary layer thickness ($\sqrt{2\mu/\omega}$) and is expressed as follows:

$$k_h = \frac{\sqrt{2}d}{2}\bigg/\sqrt{\frac{2\mu}{\omega}} = \frac{d}{2}\sqrt{\frac{\omega}{\mu}} \tag{9}$$

where $\mu = \eta/\rho_0$ is the kinematic viscosity.

The displacement of the panel is expanded in terms of the structural modal shape $\phi_n$ as follows:

$$w(x,y,t) = \sum_{n=1}^{\infty}\varphi_n(x,y)W_n e^{i\omega t} \tag{10}$$

The solutions under the simply supported boundary condition are $\varphi_{m,n}(x,y) = \sin\left(\frac{m\pi x}{L_x}\right)\cdot\sin\left(\frac{n\pi y}{L_y}\right)$.

The pressure in the cavities is expanded in terms of acoustic modal shape $\psi_n$ as follows:

$$p(x,y,z,t) = \sum_{n=0}^{\infty}\psi_n(x,y,z)P_n e^{i\omega t} \tag{11}$$

where $W_n$ and $P_n$ denote the amplitude of the modal displacement response and the modal pressure response, respectively. The solutions are $\psi(x,y,z) = \cos\left(\frac{a\pi x}{L_x}\right)\cos\left(\frac{b\pi y}{L_y}\right)\cos\left(\frac{d\pi z}{L_z}\right)$.

The structural modal shapes satisfy the following characteristic equations:

$$D^{plate}\nabla^4\varphi_n^{plate}(x,y) - \left(\omega_n^{plate}\right)^2 m^{plate}\varphi_n^{plate}(x,y) = 0 \tag{12}$$

where the superscript "plate" represents the IN, MPP, and RE. The acoustic modal shapes satisfy the following characteristic equations:

$$c^2\nabla^2\psi_n^{cavity}(x,y,z) + \left(\omega_n^{cavity}\right)^2\psi_n^{cavity}(x,y,z) = 0 \tag{13}$$

where the superscript "cavity" represents either cavity I or cavity II. In equations (12) and (13), $\omega_n$ denotes the $n$th structural or acoustic natural angular frequency. We substitute Eqs. (6)–(13) into (1)–(5) and apply the orthogonality properties of the modal shapes to obtain the following equations:

For the incidence panel:

$$\left[\left(\omega_j^{IN}\right)^2 - \omega^2 + 2i\zeta\omega\omega_j^{IN}\right]M_j^{IN}W_j^{IN} + \sum_{n=1}^{\infty}C_{nj}^{IN-I}P_n^I = P_j^{IN} \tag{14}$$

For the MPP:

$$\left[\left(\omega_j^{MPP}\right)^2 - \omega^2 + 2i\zeta\omega\omega_j^{MPP}\right]M_j^{MPP}W_j^{MPP} - \sum_{n=0}^{\infty}C_{jn}^{MPP-I}P_n^I + \sum_{n=1}^{\infty}C_{jn}^{MPP-II}P_n^{II} = 0 \tag{15}$$

For the radiation panel:

$$\left[\left(\omega_j^{RE}\right)^2 - \omega^2 + 2i\zeta\omega\omega_j^{RE}\right]M_j^{RE}W_j^{RE} - \sum_{n=1}^{\infty}P_n^{II}C_{jn}^{RE\text{-}II} = 0 \tag{16}$$

For cavity I:

$$G_j^I P_j^I - \sum_{n=1}^{\infty}C_{jn}^{I\text{-}IN}W_n^{IN} + (1-\sigma)\sum_{n=1}^{\infty}C_{jn}^{I\text{-}MPP}W_n^{MPP} + \frac{i\sigma}{\omega Z^{MPP}}\sum_{n=1}^{\infty}C_{jn}^{I\text{-}II}P_n^{II} = 0 \tag{17}$$

For cavity II:

$$G_j^{II} P_j^{II} - (1-\sigma)\sum_{n=1}^{\infty}W_n^{MPP}C_{jn}^{II\text{-}MPP} + \frac{i\sigma}{\omega Z^{MPP}}\sum_{n=1}^{\infty}P_n^I C_{jn}^{II\text{-}I} + \sum_{n=1}^{\infty}W_n^{RE}C_{jn}^{II\text{-}RE} = 0 \tag{18}$$

where

$$M_j^{plate} = m\iint_S \left[\varphi_j^{plate}(x,y)\right]^2 ds,$$

$$p_j^{IN} = 2\iint_{S^{IN}} p_{inc}(x,y,z^{IN})\varphi_j^{IN}(x,y) ds,$$

$$G_j^{cavity} = \frac{\omega^2 - \left(\omega_j^{cavity}\right)^2 + 2i\xi\omega\omega_j^{cavity}}{\omega^2\rho_0 c^2}N_j^{cavity} - \frac{i\sigma}{\omega Z^{MPP}}N_j^{cavity\text{-}z^{MPP}},$$

$$N_j^{cavity} = \iiint_\Omega \left[\psi_j^{cavity}(x,y,z)\right]^2 d\Omega,$$

$$N_j^{cavity\text{-}z^{MPP}} = \iint_{S^{MPP}} \left[\psi_j^{cavity}(x,y,z^{MPP})\right]^2 ds,$$

$$C_{jn}^{plate\text{-}cavity} = \iint_{S^{plate}} \varphi_j^{plate}(x,y)\psi_n^{cavity}(x,y,z^{plate}) ds,$$

$$C_{jn}^{cavity-plate} = \iint_{S^{plate}} \varphi_n^{plate}(x,y)\psi_j^{cavity}(x,y,z^{plate}) ds.$$

$$C_{jn}^{cavity1-cavity2} = \iint_{S^{MPP}} \psi_j^{cavity1}(x,y,z^{MPP})\psi_n^{cavity2}(x,y,z^{MPP}) ds$$

In the above equations, $\xi_j$ denotes the modal damping ratio of the $j$th mode, $\zeta_j$ denotes the $j$th modal factor of the cavity, $M_j$ denotes the $j$th modal mass of the panels, and $P_j^{IN}$ denotes the amplitude of the $j$th modal pressure. Additionally, $C_{jn}$ denotes the modal coupling coefficient characterizing the coupling strength between the $j$th structural or acoustic mode and $n$th acoustic mode. Equations (14)–(18) constitute a system of linear equations that consists of a series of equations with five subsystems, including three plates and two air cavities. The modality of each subsystem is truncated, and the number of truncations is represented by $T_x$, where $x$ is the number of truncated modalities. The modalities are combined into a group of linear equations as follows:

$$\begin{bmatrix} G^{IN} & C^{IN-I} & 0 & 0 & 0 \\ -C^{I-IN} & G^{I} & (1-\sigma)C^{I-MPP} & \dfrac{i\sigma}{\omega Z^{MPP}}C^{I-II} & 0 \\ 0 & -C^{MPP-I} & G^{MPP} & C^{MPP-II} & 0 \\ 0 & \dfrac{i\sigma}{\omega Z^{MPP}}C^{II-I} & -(1-\sigma)C^{II-MPP} & G^{II} & C^{II-RE} \\ 0 & 0 & 0 & -C^{RE-II} & G^{RE} \end{bmatrix} \begin{bmatrix} W^{IN} \\ P^{I} \\ W^{MPP} \\ P^{II} \\ W^{RE} \end{bmatrix} = \begin{bmatrix} P^{IN}_{n\times 1} \\ 0 \\ 0 \\ 0 \\ 0 \end{bmatrix} \quad (19)$$

where

$$G = \begin{bmatrix} G_1 & 0 & \cdots & 0 \\ 0 & G_2 & & \vdots \\ \vdots & & \ddots & 0 \\ 0 & \cdots & 0 & G_{T_n} \end{bmatrix}, \quad C = \begin{bmatrix} C_{11} & C_{12} & \cdots & C_{1T_m} \\ C_{21} & \ddots & & \vdots \\ \vdots & & \ddots & C_{T_{n-1}T_m} \\ C_{T_n 1} & \cdots & C_{T_n T_{m-1}} & C_{T_n T_m} \end{bmatrix}, \quad G_j^{plate} = \left[\left(\omega_j^{plate}\right)^2 - \omega^2\right] M_j^{IN}.$$

Equation (19) consists of a set of linear equations with the modal pressure of the cavities and modal displacement of the panels as unknowns. The equation is solved, and the pressure response inside the cavities and normal vibration displacement of the coupled panels are obtained.

The coupled algebraic equation (19) is solved using MATLAB software. The solution provides all the unknown modal amplitudes for the plates and cavities. They are used to calculate the transmission loss $TL$ of the system. To obtain the $TL$ of the system, it is necessary to calculate the incident and transmitted sound powers. Incident power $\Pi_{inc}$ is defined as follows:

$$\Pi_{inc} = \frac{1}{2}\text{Re}\iint_{S^{IN}} p_{inc} v_{inc}^* dS \quad (20)$$

where $S^{IN}$ denotes the area of the incident plate, $v_{inc}$ denotes the incident velocity, and the superscript * denotes complex conjugate. The incident plate is excited by a plane wave with incidence angle θ, and thus, equation (20) is expressed as follows (Bravo, Maury, & Pinhède, 2012):

$$\Pi_{inc} = \frac{|p_{inc}|^2 \cos(\theta)}{2\rho_0 c} S^{IN} \quad (21)$$

The sound power transmitted by the radiation plate is given as follows:

$$\Pi_{trans} = \frac{1}{2}\text{Re}\iint_{S^{RE}} p_{trans} \left(v^{RE}\right)^* dS \quad (22)$$

where $p_{trans}$ denotes the sound pressure radiated by the radiation panel on its surface. It is expressed in terms of $v^{RE}$ using Rayleigh's integral representation of radiated pressure. The $TL$ of the system is obtained as follows:

$$TL = -10\log_{10}(\tau) \quad (23)$$

where $\tau$ denotes the power transmission ratio that is the ratio of transmitted power $\Pi_{trans}$ to incident power $\Pi_{inc}$. This is expressed as follows:

$$\tau = \frac{\Pi_{trans}}{\Pi_{inc}} \quad (24)$$

## 3. EXPERIMENTAL VALIDATION

The model proposed in the preceding section is validated based on the aforementioned assumptions and simplifications. Therefore, to confirm the assumptions and simplifications and investigate the possibility of the practical applications by employing MPP and the air-cavity-subdivision technique, experiments were performed using a rigid box using heavy and rigid boards. The experimental set-up is shown in Figure 2. High-density boards with a thickness of up to 50 mm, density of 900 kg/m$^3$, elastic modulus of approximately 500 MPa, and Poisson's ratio of 0.25 are employed to build the semi-reverberation tank. The rigid box is considered as the object mostly isolated from exterior sound. A speaker with a diameter of approximately 30 cm is placed at a distance of 3.5 m from the incident plate to simulate the incident plane waves. The double panel and the MPP are composed of aluminum with a size of 0.6 m × 0.8 m and their Young's modulus is 6.9 × 10$^{10}$ N/m$^2$, density is 2700 kg/m$^3$, Poisson's ratio is 0.33, and structural damping ratio is 0.01. The MPP is 0.25 mm thick with holes of 0.2 mm diameter and perforation ratio of 1.3%. The incident plate is at a distance of 0.2 m from the radiation panel and 3 mm thick. The radiation panel is 4 mm thick. The size of the rigid box is 0.6 m × 0.8 m × 1 m. The structural and materials parameters are listed in Table 1 below.

As the sound field in the transmission side is now confined within a rigid box of finite size, the formula for sound power in semi-infinite space in Eq. (22) is now replaced by the averaged sound energy density in the box, given by the equation below.

$$\Pi_{trans} = \frac{\iiint_V |p(x,y,z)|^2 dV}{L_x L_y L_z} \tag{25}$$

In the actual experiments, the sound field at 24 positions inside the box was averaged.

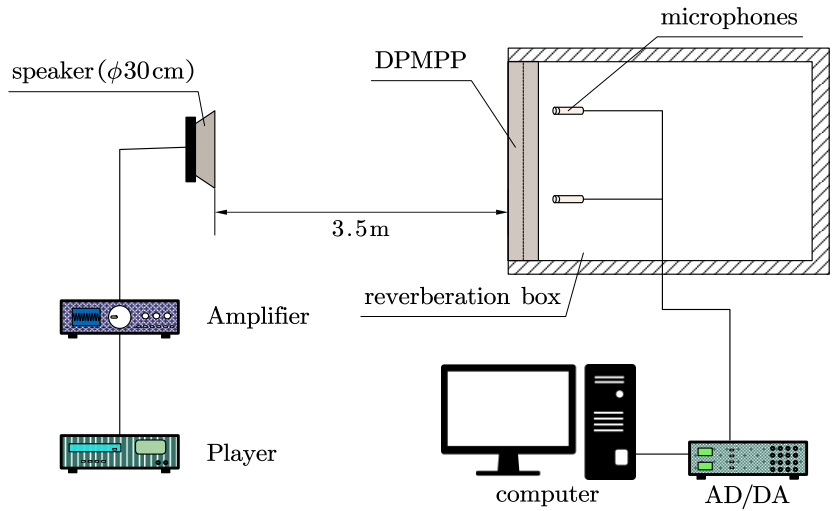

Figure 2 Experimental set-up to measure the transmission loss using a rigid box to shield external noise

Figure 3 compares the measured STL spectrum with the one obtained from the theory. The solid line in the figure is the calculated spectrum, whereas the dashed line denotes the spectrum obtained from the experimental tests. The experimental data agree well with the theoretical results and suggest that the coupled model is effective for predicting the general acoustical properties of the DPMPP. At frequencies below 250 Hz, the theoretical curve matches well with the experimental. In particular, the experimental frequency of the first resonant mode, which is manifested as a STL dip around 170 Hz, is in good agreement with the one predicted by theory. The STL peak near 140 Hz is due to the anti-resonance of the plates with surface averaged displacement close to zero [30]. The theory not only predicted correctly the frequency correctly, but also the magnitude of the STL. The second and the third resonances near 200 Hz and 230 Hz also match well with the experimental STL. At higher frequencies, the distribution of the resonant modes becomes denser, as expected, and the details of the theoretical spectrum start to deviate from the experimental one, because the box resonance effect was not fully considered in the theoretical model. In the experiment spectrum, there are more TL dips and peaks, which are likely due to the resonant modes of the reverberation box. The theoretical spectrum curve traces nearly the middle values of these peaks and dips, providing good estimation of the order of magnitudes in STL. This is more clearly shown in the 1/3 oct spectrum, where the overall agreement between theory and experiment is good.

Table 1 Geometrical parameters of the calculated DPMPP

| Parameters | Value |
| --- | --- |
| Thickness of the incidence panel | 3 mm |
| Thickness of the radiation panel | 4 mm |
| Size of the incidence panel | 0.6 m × 0.8 m |
| Size of the radiation panel | 0.6 m × 0.8 m |
| Size of the MPP | 0.6 m × 0.8 m |
| Distance between the incidence panel and radiation panel | 0.2 m |

Figure 4 shows the theoretical and experimental STL of the DPMPP. Like in the double panel case, the overall agreement between theory and experiment is good. In particular, the main features discussed for the double panel case, such as the low frequency resonance and anti-resonance, can also be clearly identified in both the experimental and theoretical STL. For more clear comparison of the effect of the introduction of the MPP, the experimental STL for the double panel and for the DPMPP are duplicated in Fig. 5(a), while the corresponding theoretical ones are shown in Fig. 5(b). As seen in Fig. 5(a), the experimental STL peak due to the first anti-resonance around 170 Hz for the DLMPP was reduced as compared to the double panel case. This is expected as the additional dissipation introduced by the MPP would reduce the anti-resonance strength by introducing a larger imaginary part in the surface average vibration field of the transmission plate. The STL dips at low frequency around 170 Hz, however, was not significantly reduced as the effect of dissipation by the MPP in the low frequency regime has little effect on the resonance-induced SLT dip. The dissipation of MPP increases with increasing frequency, which is expected due to the viscous nature of the MPP dissipation. The depths of the STL dips of the double panel are much reduced by the DPMPP starting from 200 Hz. Around 1000 Hz the improvement of STL of the DPMPP over the double panel becomes more significant. For example, near 1000 Hz the STL is enhanced by nearly 10 dB.

The above trend observed by experiments is reproduced correctly by the theory. As seen in Fig. 5(b), below 400 Hz, slight improvements in STL dips are achieved by the DPMPP over the double panel. The large depression of the double panel STL between 700 Hz and 900 Hz is almost completely leveled in the DPMPP STL. Above 1000 Hz, the improvement in STL by the DPMPP is also clearly visible. Summarizing the results presented so far, the advantage of sound insulation capability of DPMPP over that of the double panel is clearly evident as supported by both experimental and theoretical investigations.

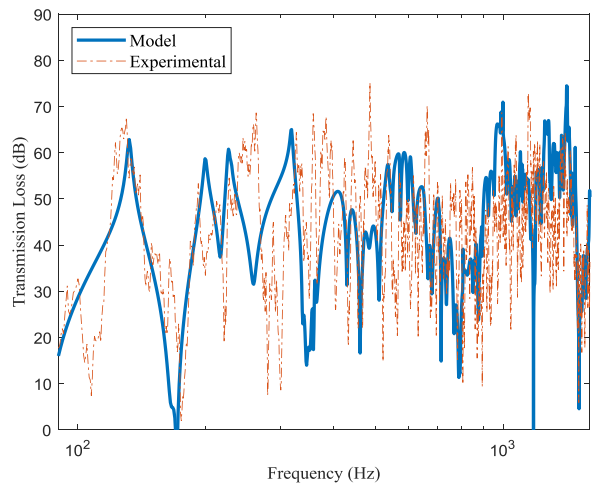

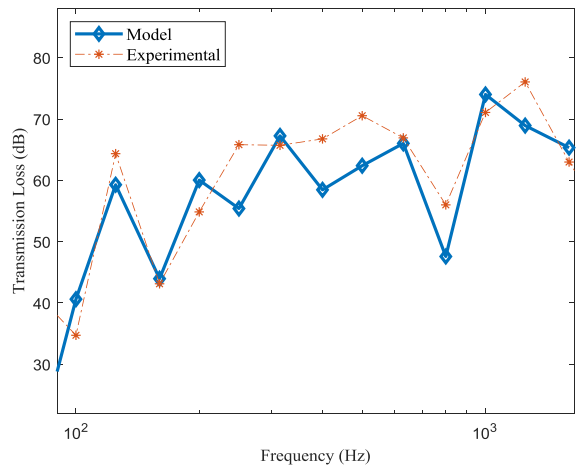

Figure 3 Double Plate theoretical results vs experimental results in (a) high resolution spectrum (b) 1/3oct spectrum.

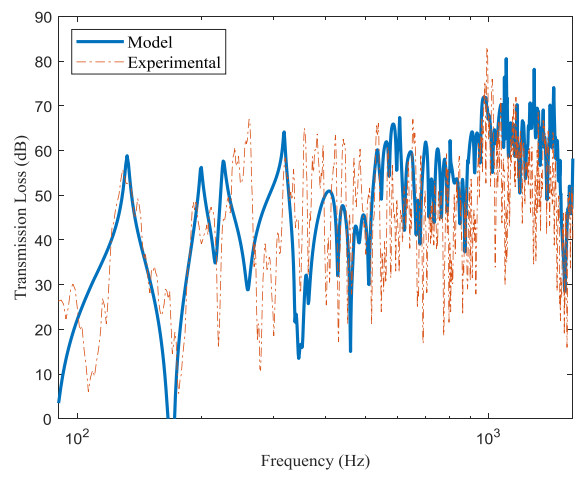

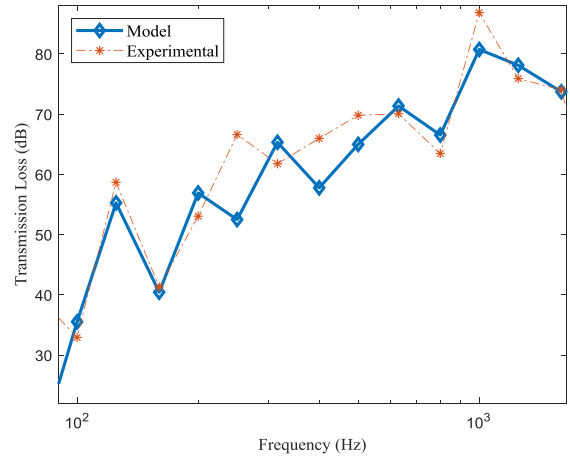

Figure 4 DPMPP theoretical results vs experimental results in (a) high resolution spectrum (b) 1/3oct spectrum.

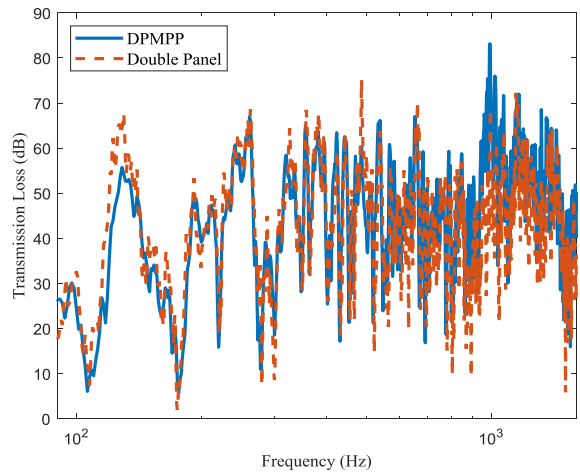

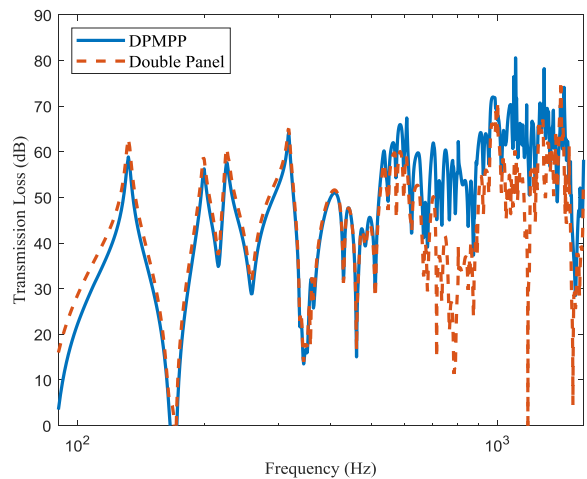

Figure 5 (a) High resolution experimental DPMPP STL (solid curve) and double panel STL (dashed curve); (b) Theoretical DPMPP STL (solid curve) and double panel STL (dashed curve).

## 4. FURTHER THEORETICAL INVESTIGATIONS AND DISCUSSIONS

Having established the advantage of MPP in enhancing STL in the double panel structure, we now numerically investigate further the effects of the MPP on the STL with different MPP positions and structural parameters.

An MPP with a thickness of 0.25 mm is inserted in the double panel. The other geometrical parameters of these panels are listed in Table 1. The MPP position $\zeta$ is defined as the ratio of the distance between the incidence plate and the radiation plate over that between the incidence plate and MPP as follows:

$$\zeta = \frac{z_1}{z_3} \times 100\% \quad (26)$$

When $\zeta$ is equal to 50%, the MPP is located at equal distance to the incident and radiation panels, which is the case in the previous chapter. When the value is less than 50%, the MPP is closer to the incident panel. When the value exceeds 50%, the MPP is closer to the radiation panel. Due to transmission reciprocity, only the variation of $\zeta$ between 0 and 50% is considered here. The cases for a normal plate of the same thickness as the MPP in replacement of the MPP in the cavity (triple-panel) are also studied for comparison.

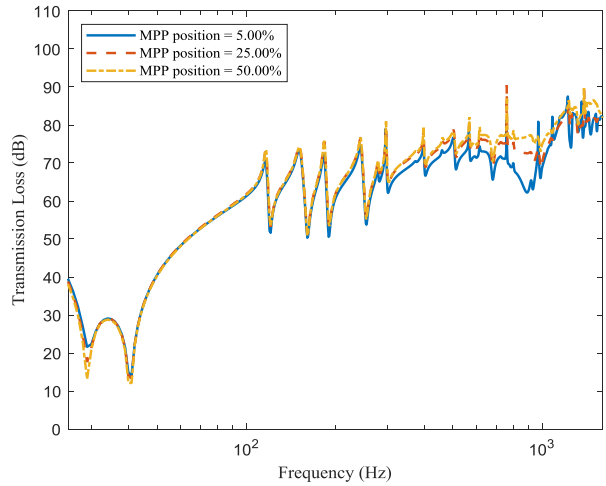

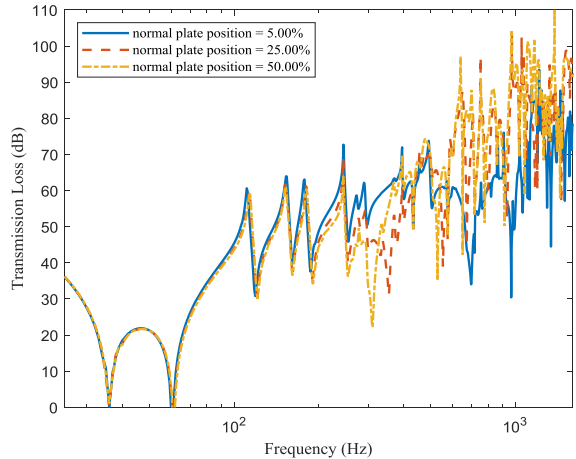

Figure 3 (a) STL of the DPMPP at different positions; (b) STL of normal plate at different positions.

Figure 3 compares the STL of a DPMPPs at different positions. The sound waves passing through the MPP is dissipated by the viscous and thermal dissipations inside the pores of the MPP. It is observed that when the MPP is closer to the incident plate, the first-order resonance of the system exhibits a significant reduction in comparison with the case where MPP is mid-way between the two plates, resulting in about 5.6 dB in STL, while in the middle and high frequencies the STL favors $\zeta = 50\%$. The resonance of the two cavities significantly contributes to the first-order resonance of the system. When $\zeta$ is close to 0 or 1, the difference in the thickness of the two cavities reaches a maximum. This also makes the fundamental frequencies of the two cavities to be well separated, causing weakening of the first-order resonance of the entire system. At $\zeta = 50\%$, the dimensions of the two cavities are exactly identical, causing the two cavities to resonate through the micro-pores. This results in a significant reduction in the STL at the fundamental frequency. However, the resonance effect between the two cavities causes the MPP to dissipate more acoustic energy at higher frequencies, and this in turn increases the overall STL. To obtain a higher total STL, $\zeta$ should be near 50%. Hence, further examinations of the perforation parameters are conducted only with the fixed value of $\zeta = 50\%$, i.e., with the MPP located in the middle of the incidence and radiation plates. The effect of the perforation ratio on the MPP, the hole diameter, and the MPP thickness are discussed in the following sections.

The cases where a normal plate of the same thickness as the MPP is in place for the MPP are also investigated, and the results are shown in Fig. 6(b). It is seen that at the first two resonant frequencies of 110 Hz and 120 Hz the STL is nearly zero. The improvement in STL by the MPP $\zeta = 5\%$ is 22 dB at 100 Hz, and 15 dB at 120 Hz. The MPP therefore is a better choice than a normal plate in the low frequency, because the MPP reduces the resonant strength. At $\zeta = 50\%$, the MPP still improves the STL by 14 dB at 110 Hz, and 13 dB at 120 Hz.

Here we discusses the effect of the perforation ratio on the STL of the DPMPP with $\zeta$ fixed at 50%. An MPP or a normal plate of thickness 0.25 mm is inserted in the middle of the double panel. The STL spectra of a triple-panel and with a DPMPP structure with various MPP perforation ratios are shown in Fig. 7. The other parameters of these panels are listed in Table 1. It is observed that when compared with the classical triple panels, the DPMPP partitions exhibit a larger STL at low frequencies. The mass–air–mass resonances of the classical triple panels occurs at approximately 38 Hz and 62 Hz. The STL is approximately 0 dB, whereas the resonance of the DPMPPs occurs at approximately 40 Hz. Because of the presence of the MPP, the resonance at approximately 62 Hz is removed, and the SLT of the DPMPP is significantly increased at low frequencies. The STL at 1.3% perforation ratio is the highest at most frequencies below 600 Hz, only to be surpassed by the STL with lower perforation MPP cases in the higher frequencies.

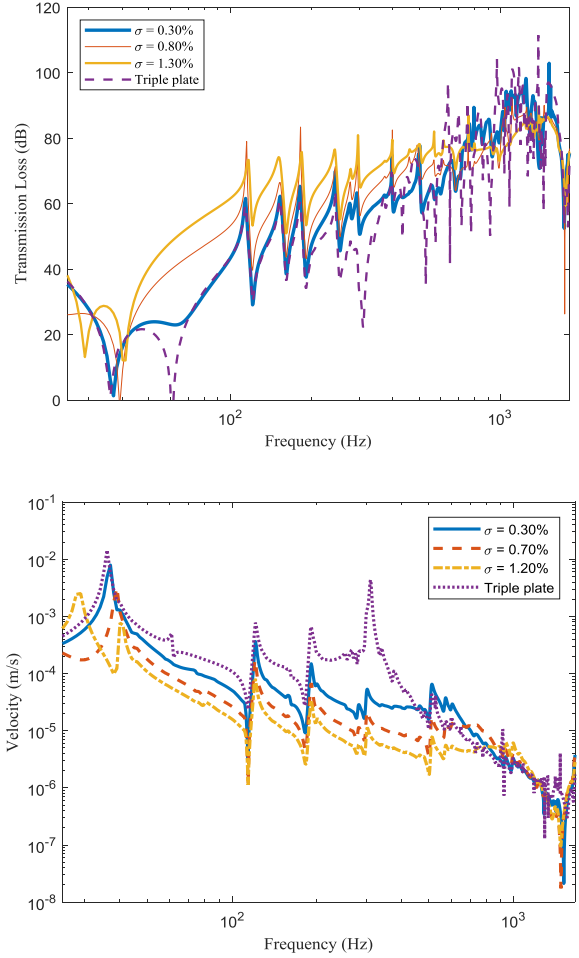

Figure 7 (a) STL and (b) Normal average velocity of the triple-panel and the DPMPP at different MPP perforation ratios.

As our theoretical model incorporates the motion of the MPP, we plot the surface averaged normal velocity of the MPP with different perforation ratios and that of the normal plate in the triple-panel structure. It can be seen that the velocity and therefore the displacement of the MPP with lower perforation ratio is higher than that of the higher ratio, and the normal plate is significantly higher in the 200 Hz to 500Hz range. This is because the holes in the MPP release the pressure difference at the front and back surface of the MPP, the more so for higher perforation ratio, while the rigidity of the MPP decreases with increase perforation ratio. At frequencies above 1000 Hz, the viscous friction made the holes less and less transparent for air flow, while its rigidity is always less than the solid plate, so that the displacement becomes comparable to the solid plate. One may expect the MPP displacement to exceed that of the solid plate at even higher frequencies which are outside our investigation range.

Next we discusses the effect of the hole diameter of the MPP on the TL of DPMPP. The MPP is located in the middle position between the incident and the radiation plates. The perforation rate and the position of the MPP were kept constant, while changing the micro-pore diameter. As shown in Fig. 8, an increase in the hole diameter increases the STL at low frequencies but decreases it at intermediate to high frequencies. When the perforation ratio is fixed, the air-frame interfacial area inside the perforated holes increases when the hole diameter is reduced. The reduced air-frame interface area increases the acoustic resistance and consequently increases the STL of the DPMPP. An increase in the diameter of the small hole significantly decreases its acoustic resistance. The STL exhibits further improvements when the frequency is further decreased. Additionally, the increase in the rate of each frequency is essentially identical (Putra & Thompson, 2010). Therefore, it is determined that the decrease in the acoustic resistance of a micro-perforated plate causes the increase in the structure acoustic propagation loss. Figure 8 also presents the theoretical results obtained using a normal plate with the same parameters albeit without micro-pores, as opposed to the MPP. As shown in Fig. 8, in a manner similar to the results in the previous section, the second-order resonance at approximately 60 Hz disappears after the middle plate is replaced by MPP. An increase in the hole diameter continues to decrease the resonance at approximately 38 Hz, and it split into two lower dips. At medium and high frequencies, the change in the MPP hole diameter does not significantly affect the STL.

Similar to the cases in the previous section, the averaged normal displacement of the MPP in the DPMPP is significantly smaller than the solid plate in the triple-panel structure. The reasons for such phenomenon are much the same as the ones in the previous section.

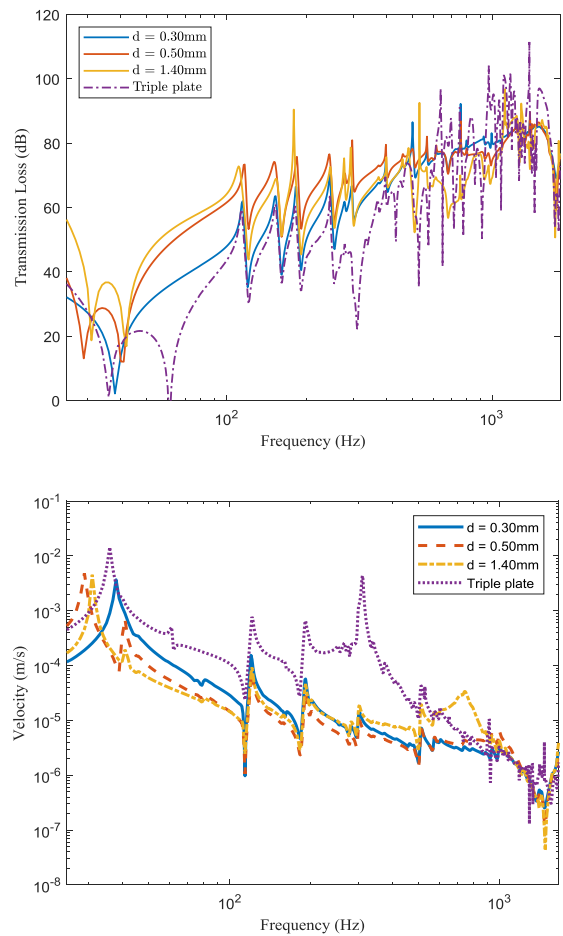

Figure 8 (a) STL and (b) Averaged normal velocity comparison of the triple-panel and the DPMPP structure at different hole diameters.

## 5. Conclusions

In the work, the sound transmission loss improvement by the introduction of a micro perforated plate in a cavity formed by two normal plates is systematically investigated. An analytical model that takes into account of the full coupling of the major components of the double-panel micro perforation plate structure, namely the two panels of the air cavity, the air in the cavity, and the perforation plate, has been developed, and tested by experimental results. The experiment verifies the validity and reliability of the theoretical model using a reverberation box. The improvement in the transmission loss at low frequencies is significant when compared to the cases without the MPP, and when the MPP is replaced by a normal solid plate of the same thickness as the MPP. The major findings are summarized below: 1) When the MPP is placed in the middle of the cavity, the STL in the medium and high frequencies is improved, and the first-order resonance of the structure is suppress when it is set on one side. 2) The STL of the structure at low frequencies increases with the increase of the perforation rate of the MPP and the diameter of the small hole, whereas the effects of the perforation rate and diameter of the small hole are not evident at high frequencies. 3) The displacement of the MPP skeleton can be ignored in the frequency range below about 600 Hz, while it is comparable to a solid plate of the same thickness in the high frequency range. The results obtained in the present study can aid researchers in designing a high-performance multi-layer structure that uses an MPP to improve the low-frequency sound insulation performance of conventional multi-plate structures and acoustic metamaterials.